\documentclass[a4paper]{jpconf}
\usepackage{graphicx}
\usepackage{amssymb}
\usepackage{amsmath}
\usepackage{latexsym}
\def\vev#1{\left\langle #1\right\rangle}
\begin{document}
\title{Discrete dark matter mechanism}

\author{E. Peinado}

\address{ AHEP Group, Institut de F\'{\i}sica Corpuscular --
  C.S.I.C./Universitat de Val{\`e}ncia \\
  Edificio Institutos de Paterna, Apt 22085, E--46071 Valencia, Spain}

\ead{epeinado@ific.uv.es}

\begin{abstract}
We present the Discrete Dark Matter mechanism (DDM) which consist on the stability of the dark matter from a spontaneous breaking of a flavor symmetry group into one of its subgroups. Here we present the inclusion of the quarks on this mechanism promoting the quarks to transform non-trivial under the flavor group.\end{abstract}
\section{Introduction}
Non-baryonic Dark Matter (DM) is one of the most compelling 
problems of modern cosmology.
Despite the fact that its existence is well established by cosmological and astrophysical
probes, its nature remains elusive. 
Still, observations can constraint the properties of dark matter and give some hints about its identity. One of the fundamental
requirements for a viable dark matter candidate is the stability over cosmological times. 

This stability could be due to the existence of
a symmetry protecting or suppressing its decay. It has been shown recently that such symmetry can be related
to the flavor structure of the Standard Model~\cite{Hirsch:2010ru}. The model proposed in ~\cite{Hirsch:2010ru} is based on a $A_4$ symmetry with four $SU(2)$ Higgs doublets. After the electroweak symmetry breaking, the $A_4$ (even permutation of four objects) group is spontaneously broken into a $Z_2$ subgroup which is responsible for the DM stability. The leptonic sector is also extended. It consist in four right handed neutrinos and the light neutrino masses are generated through the type-I seesaw mechanism and obey an inverted hierarchy mass spectrum with $m_{\nu_3}=0$ and vanishing reactor angle $\theta_{13}=0$. 

We have also been consider models with a different matter content for the right handed neutrinos with the same DM stability mechanism but with different neutrino phenomenology~\cite{Meloni:2010sk} or a model based on the dihedral group $D_4$ where the some flavor changing neutral currents are present and constraint the DM sector~\cite{Meloni:2011cc}. There are some models based on flavor symmetries but with decaying DM have also been considered, see for instance~\cite{Daikoku:2010ew,arXiv:1011.5753}. For a model with non-abelian flavor symmetries with stable DM but where the the DM couples to some right handed neutrinos in a similar way to our model\footnote{In this model the DM stability is due to an extra $Z_2$ where the DM and some right handed neutrinos are charged under this symmetry.} see~\cite{Eby:2011qa}. 

\section{Discrete Dark Matter Mechanism}

The discrete dark matter (DDM) mechanism consist on the stability of the dark matter is by means a residual $Z_N$ symmetry of a spontaneously broken discrete flavor symmetry, for instance in the original the group of even permutations of four objects was used, $A_4$, and it was spontaneously broken by the electroweak symmetry breaking mechanism into a $Z_2$ sub group, $A_4\rightarrow Z_2$. 

The $A_4$ group has two generators: S, and T, which obeys the properties $S^2=T^3=(ST)^3=I$. $S$ is a $Z_2$ generator while $T$ is a $Z_3$ generator. $A_4$ has four irreducible representations, three singlets $1$, $1^\prime$, and $1^{\prime \prime}$ and one triplet. The generators of $A_4$ in the $S$-diagonal basis are
\begin{equation}
\begin{array}{lr}
\begin{array}{ccc}
1 & S=1 & T=1\\
1^\prime & S=1 & T=\omega^2\\
1^{\prime\prime} & S=1 & T=\omega
\end{array}& 3:\begin{array}{lr}
S=\left(
\begin{array}{ccc}
1&0&0\\
0&-1&0\\
0&0&-1\\
\end{array}
\right)&T=\left(
\begin{array}{ccc}
0&1&0\\
0&0&1\\
1&0&0\\
\end{array}
\right)
\end{array}
\end{array}
\label{generators}
\end{equation}
where $\omega^3=1$. 

The model consist on and extended SM where instead of have one $SU(2)$ doublet of scalar field, we have three of them transforming as a triplet under $A_4$, and four right handed neutrinos three of them transforming a triplet and the other one as a singlet of the flavor group. In table~\ref{tab1} we present the relevant quantum numbers for the matter fields.
\begin{table}[h!]
\begin{center}
\begin{tabular}{|c|c|c|c|c|c|c|c|c||c|c|}
\hline
&$\,L_e\,$&$\,L_{\mu}\,$&$\,L_{\tau}\,$&$\,\,l_{e}^c\,\,$&$\,\,l_{{\mu}}^c\,\,$&$\,\,l_{{\tau}}^c\,\,$&$N_{T}\,$&$\,N_4\,$&$\,H\,$&$\,\eta\,$\\
\hline
$SU(2)$&2&2&2&1&1&1&1&1&2&2\\
\hline
$A_4$ &$1$ &$1^\prime$&$1^{\prime \prime}$&$1$&$1^{\prime \prime}$&$1^\prime$&$3$ &$1$ &$1$&$3$\\
\hline
\end{tabular}\caption{Summary of relevant model quantum numbers}\label{tab1}
\end{center}
\end{table}
The Yukawa Lagrangian of the model is given by
\begin{eqnarray}\label{lag}
\mathcal{L}&=&y_e L_el_{_e}^c H+y_\mu L_\mu l_{_\mu}^c H+y_\tau L_\tau l_{_\tau}^c H++y_1^\nu L_e(N_T\eta)_{1}+y_2^\nu L_\mu(N_T\eta)_{1''}+\nonumber\\&&+y_3^\nu L_\tau(N_T\eta)_{1'}+y_4^\nu L_e N_4 H+ M_1 N_TN_T+M_2 N_4N_4+
\mbox{h.c.}
\end{eqnarray}
This way $H$ is responsible for quark and charged lepton masses, the latter automatically diagonal\footnote{For quark mixing angles generated through higher dimension operators see reference ~\cite{deAdelhartToorop:2011ad}.}. Neutrino masses arise from $H$ and $\eta$. With the vacuum alignment for the scalar fields~\cite{Hirsch:2010ru} is 
\begin{equation}
  \vev{ H^0}=v_h\ne 0,~~~~ \vev{ \eta^0_1}=v_\eta \ne 0~~~~
\vev{ \eta^0_{2,3}}=0\,,
\end{equation}
which means the vev alignment for the $A_4$ triplet of the form $\vev{ \eta} \sim (1,0,0)$. This alignment is invariant under the $S$ generator\footnote{$H_s$ is in the $1$ representation of $A_4$ and its vev also respect the generator $S$.}, see eq. (\ref{generators}), therefore the minimum of the potential breaks spontaneously $A_4$ into a $Z_2$ subgroup generated by $S$. All the fields in the model singlets under $A_4$ are even under the residual $Z_2$, the triplets transform as:
\begin{equation}\label{residualZ2}
\begin{array}{lcrlcr}
N_1 &\to& +N_1\,,\quad& \eta_1 &\to& +\eta_1 \\   
N_2 &\to& -N_2\,,\quad& \eta_2 &\to& -\eta_2 \\   
N_3 &\to& -N_3\,,\quad& \eta_3 &\to& -\eta_3.  
\end{array}
\end{equation}
The DM candidate is the lightest particle charged under $Z_2$ i. e. the lightest combination of the scalars $\eta_2$ and $\eta_3$, which we will denote
generically by $\eta_{DM}$. We list below all interactions of $\eta_{DM}$:
\begin{enumerate}
\item Yukawa interactions
\begin{equation}
\begin{array}{l}
\eta_{_{DM}}\, \overline{\nu}_i   N_{2,3}\,,
\end{array}
\end{equation}
where $i=e,\,\mu,\,\tau$.
\item Higgs-Vector boson couplings
\begin{equation}\label{eq:gint}
\begin{array}{l}
\eta_{_{DM}}\eta_{_{DM}} ZZ\,,\quad \eta_{_{DM}}\eta_{_{DM}} WW\,,\quad
\eta_{_{DM}}\eta_{2,3}^{\pm} W^\pm Z\,,\quad \eta_{_{DM}}\eta_{2,3}^{\pm} W^\pm \,,
\eta_{_{DM}}A_{2,3} Z \,.
\end{array}
\end{equation}
\item Scalar interactions from the Higgs potential:
\begin{equation}\label{eq:Pint}
\begin{array}{l}
\eta_{_{DM}}\, A_1 A_2 h\,,\quad \eta_{_{DM}}\, A_1 A_3 h_1\,,\quad 
\eta_{_{DM}}\, A_1 A_2 h_1\,,\quad \eta_{_{DM}}\, A_1 A_3 h\,,\\ 
\eta_{_{DM}}\, A_2 A_3 h_3\,,\quad\eta_{_{DM}}\, h_1 h_3 h\,\quad 
\eta_{_{DM}}\eta_{_{DM}} hh\,,\quad\eta_{_{DM}}\eta_{_{DM}} h_1h_1\,.
\end{array}
\end{equation}
\end{enumerate}
After electroweak symmetry breaking, the vevs the Higgs fields acquire vacuum expectation values, $v_h$ and $v_\eta$ for the singlet and the first component of the triplet respectively, additional terms are obtained from those in Eq.~(\ref{eq:Pint}) by replacing $ h\to v_h$ and $h_1\to v_\eta$. The dark matter phenomenology of this specific model has been studied in detail in~\cite{Boucenna:2011tj}. 
\section{A Discrete Dark Matter model for quarks and leptons}

We where looking for a group $G$ that contains at least two irreducible
representations of dimension larger than one, namely $r_a$ and $r_b$. We also require that all the components of the irreducible representation $r_a$ transform trivially under an abelian subgroup of $G\supset Z_N$
(with $N=2,3$) while at least one component of the irreducible representation $r_b$ is charged with under $Z_N$. The stability of the lightest component of the matter fields transforming as $r_b$ is guaranteed by $Z_N$ giving a potential~\footnote{Other requirements must fulfilled in order to have a viable DM
  candidate, such as neutrality, correct relic abundance, and
  consistency with constraints from DM search experiments.} DM
candidate.

The smallest group with this property we found is $\Delta(54)$, isomorphic to $(Z_3\times Z_3) \rtimes S_3$.  This group contains four triplet irreducible representations, ${\bf 3_{1,2,3,4}}$, in addition, $\Delta(54)$ contains four
different doublets ${\bf 2_{1,2,3,4}}$ and two singlet
irreducible representations, ${\bf 1_{\pm}}$.  The product rules for the doublets are as follows:
\begin{itemize}
\item The product of two equal doublets
\begin{equation}{\bf 2_k}\times {\bf 2_k}= {\bf 1_+}+{\bf 1_-}+{\bf 2_k}\end{equation} 
\item The product of two different doublets give us the other two doublets, for instance:
\begin{equation}{\bf 2_1}\times {\bf 2_2}={\bf 2_3}+{\bf 2_4}\end{equation}.
\end{itemize}
Of the four doublets ${\bf 2_1}$ is invariant under the $P\equiv (Z_3\times Z_3)$ subgroup of $\Delta(54)$, while the others transform non-trivially, for example
${\bf 2_3} \sim (\chi_1,\chi_2)$, which transforms as $\chi_1\,(\omega^2,\omega)$ and $\chi_2\,(\omega,\omega^2)$ respectively, where $\omega^3=1$. We can see that by taking $r_a= {\bf 2_1}$ and $r_b = {\bf 2_3}$ that $\Delta(54)$ is a perfect choice for our purpose.
\begin{table}[h]
\begin{center}
\begin{tabular}{ccccc|cccc}
\hline
\hline
& $\overline{L}_e$ & $\overline{L}_D$ & $e_R$ & $l_{D}$   
& $H$  & $\chi $ & $\eta $ &$\Delta$\\
\hline 
$SU(2)$ & $2$ & $2$ & $1$ & $1$  &$2$ & $2$ & $2$ & $3$\\
$\Delta (54)$ & $\bf 1_+$ & $\bf 2_1$ & $\bf 1_+$ & $\bf 2_1$    & $\bf 1_+$& $\bf 2_1$ & $\bf 2_3$ & $\bf 2_1$ \\
\hline
\hline
\end{tabular}
\caption{Lepton and higgs boson assignments of the model. }\label{tab2}
\end{center}
\end{table}

\begin{table}[h!]
\begin{center}
\begin{tabular}{cccccccc}
\hline
\hline
$ $ & $Q_{1,2}$ & $ Q_{3}$ & $(u_{R},c_R)$ &$t_R$ &$d_R$ & $s_R$ &$b_R$\\
\hline
$SU(2)$ & $2$ & $2$ & $1$ & $1$  &$1$ & $1$ & $1$ \\
\hline
$\Delta(54)$ & ${\bf 2_1}$& ${\bf 1_+}$& ${\bf 2_1}$& ${\bf 1_+}$& ${\bf 1_-}$& ${\bf 1_+}$& ${\bf 1_+}$\\
\hline
\hline
\end{tabular}\caption{Quark gauge and flavor representation assignments.}
\label{tab3}
\end{center}
\end{table}
In this way we can choose the scalars in the model to belong to the ${\bf 2_3}$ for instance while the active scalars to belong to the ${\bf 2_1}$~\cite{Boucenna:2012qb}. The relevant quantum numbers of the model are in Tables ~\ref{tab2} and ~\ref{tab3}. In table~\ref{tab2} $L_D \equiv (L_\mu ,L_\tau) $ and $l_D \equiv (\mu_R, \tau_R)$. There are five $SU(2)_L$ doublets of Higgs scalars: the $H$ is a singlet of $\Delta(54)$, while $\eta=(\eta_1,\eta_2) \sim {\bf 2_3}$ and
$\chi=(\chi_1,\chi_2) \sim {\bf 2_1}$ are doublets.  In order to
preserve a remnant $P$ symmetry, the doublet $\eta$ is not allowed to
take vacuum expectation value (vev).

Lets start discussing the quark sector. In Ref.~\cite{Hirsch:2010ru,Meloni:2011cc,Boucenna:2011tj,Meloni:2010sk}
quarks were considered blind under the flavor symmetry to guarantee the stability of the DM. Consequently the generation of quark mixing angles was difficult, see~\cite{deAdelhartToorop:2011ad}. 

A nice feature of our current model is that with $\Delta(54)$ we can
assign quarks to the singlet and doublet representations as shown in
table\,\ref{tab2}, in such a way that we can fit the CKM mixing
parameters.

\begin{figure}[h!]
\begin{center}
 \includegraphics[width=6.5cm]{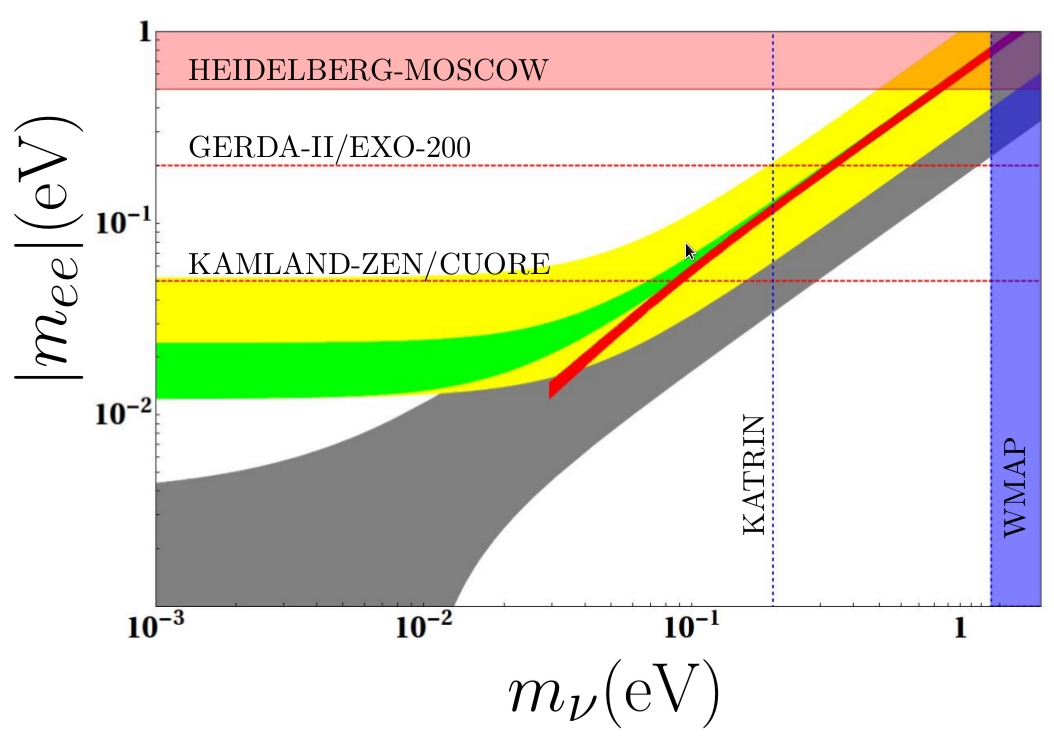}
 \caption{Effective neutrinoless double beta decay parameter $m_{ee}$
   versus the lightest neutrino mass. The thick upper and lower
   branches correspond the ``flavor-generic'' inverse (yellow) and
   normal (gray) hierarchy neutrino spectra, respectively.  The model
   predictions are indicated by the green and red (darker-shaded)
   regions, respectively. They were obtained by taking the 3$\sigma$
   band on the mass squared differences.  Only these sub-bands are
 allowed by the $\Delta (54)$ model.  For comparison we give the
 current limit and future sensitivities on
 $m_{ee}$~\cite{Schwingenheuer:2012jt,Rodejohann:2011mu} and
 $m_\nu$~\cite{Osipowicz:2001sq,Komatsu:2010fb}, respectively.
}\label{figbb}
\end{center}
\end{figure}

The resulting up- and down-type quark mass matrices in our model are
given by
\begin{equation}
M_d=
\begin{pmatrix}
ra_d & rb_d & rd_d \\
-a_d & b_d & d_d \\
0 & c_d & e_d
\end{pmatrix},\quad 
M_u=
\begin{pmatrix}
ra_u & b_u & d_u \\
b_u & a_u & rd_u \\
c_u & rc_u & e_u
\end{pmatrix}.\label{matrixquarks}
\end{equation}
where $r=\langle\chi_2\rangle/\langle\chi_1\rangle$. Note that the Higgs fields $H$ and $\chi$ are common to the lepton and the quark sectors and in particular the parameter $r$.  In order to fit of all quark masses and
mixings provided $r$ lies in the range of about $ 0.1<r<0.2$~\cite{Boucenna:2012qb}.
We turn to the leptonic sector, the charged leptons are basically equal to the up quarks under the flavor symmetry, so the charged lepton mass matrix is of the same form of the up quark mass matrix in Eq. (\ref{matrixquarks}). The neutrino masses are generated through the type-II see-saw mechanism~\cite{schechter:1980gr}. For that we include in the model an $SU_L(2)$ Higgs triplet scalar field $\Delta \sim{\bf 2_1}$.
Regarding dark matter, note that the lightest $P$-charged particle in
$\eta_{1,2}$ can play the role of ``inert'' DM~\cite{barbieri:2006dq},
as it has no direct couplings to matter. The conceptual link
between dark matter and neutrino phenomenology arises from the fact
that the DM stabilizing symmetry is a remnant of the underlying flavor
symmetry which accounts for the observed pattern of oscillations.
Choosing the vev alignment
$\vev{\Delta}\sim (1,1)$ and $\vev{\chi_1} \ne \vev{\chi_2}$,
consistent with the minimization of the scalar potential one finds
that
\begin{equation}\label{mnu}
M_\nu \propto 
\begin{pmatrix}
0 & \delta  & \delta   \\
\delta & \alpha & 0 \\
\delta  & 0  & \alpha
\end{pmatrix},
\end{equation}
where $\delta=y_a\vev{\Delta}$, $\alpha= y_b \vev{\Delta}$~\cite{Boucenna:2012qb}. 
Note that this matrix has two free parameters and gives us a neutrino mass sum rule
of the form $m_1^\nu+m_3^\nu=m_2^\nu$ in the complex plane, which has
implications for neutrinoless double beta decay~\cite{Dorame:2011eb},
as seen in Fig.~(\ref{figbb}).

\begin{figure}[h!]
\begin{center}
 \includegraphics[width=6cm]{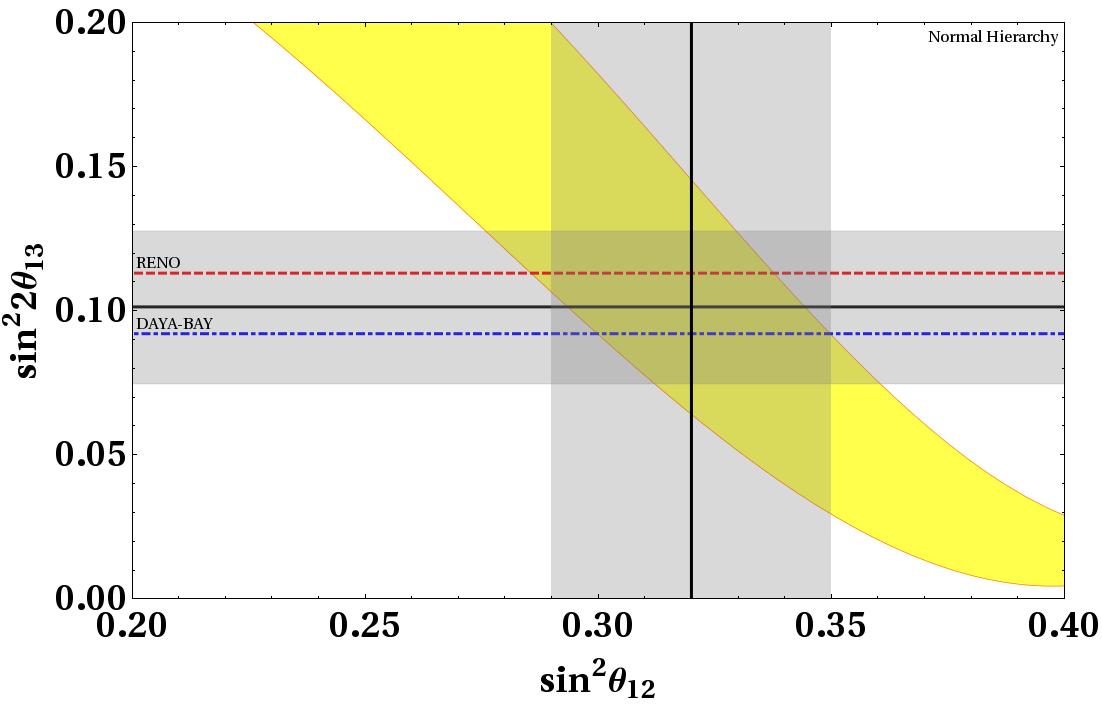}\includegraphics[width=6cm]{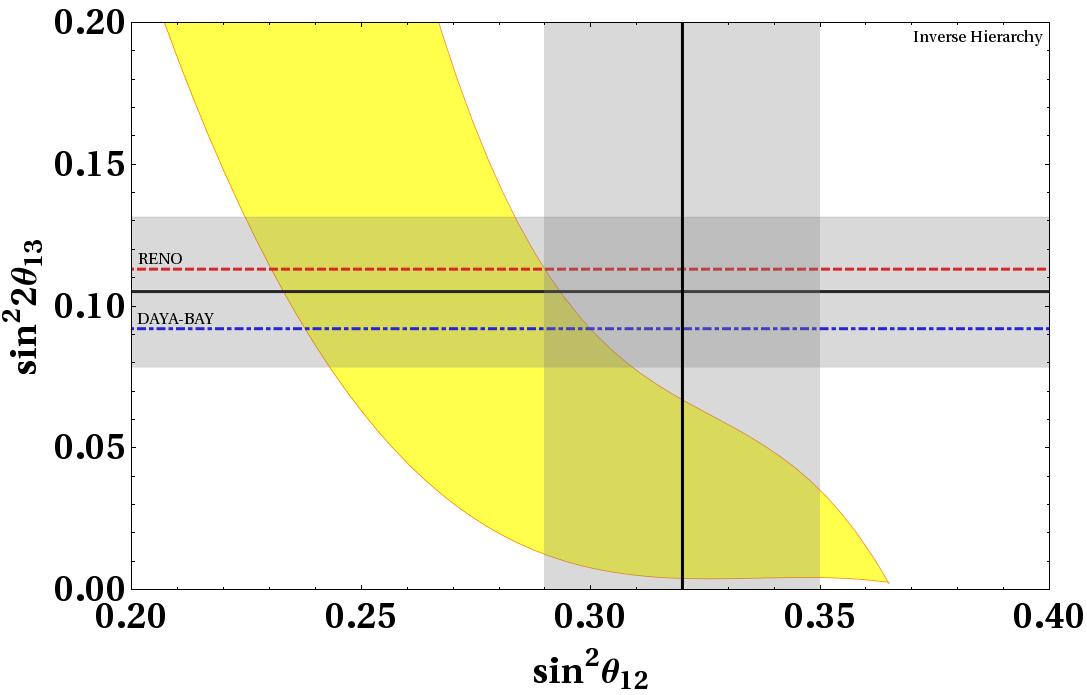}
 \caption{The left figure (right figure) is the correlation for  normal hierarchy (inverse hierarchy). The shaded (yellow) curved band gives the predicted
   correlation between solar and reactor angles when the solar and
   atmospheric squared mass splittings are varied within 2$\sigma$ for
   the normal hierarchy spectrum.  The solid (black) line gives the
   global best fit values for $\theta_{12}$ and $\theta_{13}$, along
   with the corresponding two-sigma bands, from
   Ref.~\cite{Schwetz:2011zk}. The dashed lines correspond to the
   central values of the recent published reactor
   measurements~\cite{An:2012eh,collaboration:2012nd}. Note that
   $\theta_{23}$ is also within 2$\sigma$}\label{figcn}
\end{center}
\end{figure}

We now turn to the second prediction.  For simplicity, we consider in
what follows only real parameters and we fix the intrinsic neutrino
CP--signs~\cite{Schechter:1981hw} as $\eta=diag(-,+,+)$, where $\eta$
is defined so that the CP conservation condition in the charged
current weak interaction reads $U^\star=U \eta$, $U$ being the lepton
mixing matrix. The correlations we have are presented in Fig.~\ref{figcn}.

\section{Conclusions}

We have extended DDM to include the quarks in the game. We have described how spontaneous breaking of a $\Delta(54)$
flavor symmetry can stabilize the dark matter by means of a residual
unbroken symmetry. In our scheme left-handed leptons as well as quarks
transform nontrivially under the flavor group, with neutrino masses
arising from a type-II seesaw mechanism.  We have found lower bounds
for neutrinoless double beta decay, even in the case of normal
hierarchy, as seen in Fig.~\ref{figbb}. In addition, we have
correlations between solar and reactor angles consistent with the
recent Daya-Bay and RENO reactor measurements, see Fig.~\ref{figcn}, interesting in their own right.  Direct and indirect detection prospects are similar to a generic WIMP
dark matter, as provided by multi-Higgs extensions of the SM, see, for
example, Ref.~\cite{Boucenna:2011tj}.

\section{Acknowledgments}
I thank my collaborators Sofiane Boucenna, Stefano Morisi, Yusuke Shimizu and Jose W. F. Valle for the work presented here.
This work was supported by the Spanish MICINN under grants
FPA2008-00319/FPA, FPA2011-22975 and MULTIDARK CSD2009-00064
(Consolider-Ingenio 2010 Programme), by Prometeo/2009/091 (Generalitat
Valenciana), by the EU ITN UNILHC PITN-GA-2009-237920 and by CONACyT (Mexico).

\end{document}